\documentstyle[multicol,epsfig,aps]{revtex}

\begin{document}
\title{On the possibility of
superconductivity in PrBa$_{2}$Cu$_{3}$O$_{7}$.}
\author{I.I. Mazin}
\address{Code 6691, Complex Systems Theory Branch,
Naval Research Laboratory, 4555 Overlook Ave. N.W., Washington, DC 20375}
\date{\today}
\maketitle

\begin{abstract}
Recent reports about observations of superconductivity in PrBa$_2$Cu$_3$O$_7$
raise a number of questions: (i) of various theories striving to
explain the $T_c$ suppression in Pr$_x$Y$_{1-x}$Ba$_2$Cu$_3$O$_7$,
are there any compatible with possible superconductivity in stoichiometric
PrBa$_2$Cu$_3$O$_7$? (ii) if this superconductivity is not an
experimental artifact, are the superconducting carriers (holes) of the
same character as in the other high-$T_c$
cuprates, or do they represent another electronic subsystem? (iii)
is the underlying mechanism the same as in other high-$T_c$
superconductors? I present an answer to the first two questions,
while leaving the last one open.
\end{abstract}

\begin{multicols}{2}
One of the most exciting cases of superconductivity suppression in high-$%
T_{c}$ cuprates is that of $RE_{1-x}$Pr$_{x}$Ba$_{2}$Cu$_{3}$O$_{7},$ where $%
RE$ stands for a rare earth (see Refs.\cite{rad,PM} for reviews). Even more
exciting are recent indications that conductivity and superconductivity can
be restored in pure stoichiometric PrBa$_{2}$Cu$_{3}$O$_{7}$\cite{Dow,Jap}.
This is such an unexpected results\cite{lee} that it is still not generally
accepted and further experimental confirmation is required. Nevertheless,
this fact was reported by two independent groups, and it is time now to
understand the theoretical consequences of this finding. The most 
important message, if this finding is true, is that at $x=1$, and,
presumably, at intermediate $x$'s, there are free carriers in $RE_{1-x}$Pr$_x
$Ba$_{2}$Cu$_{3}$O$_{7}$, and the suppression  of metallic conductivity at
sufficiently large $x$ must be due to localization of those carriers.
This statement effectively eliminates the possibility of hole depletion due to
hole transfer into occupied states (``four-valent Pr model''). It 
furthermore becomes highly unlikely that any kind of magnetic pair
breaking is in effect, because (1) normal state conductivity drops sharply
with doping, indicating the change of character, if not the number,
of carriers, and (2) superconductivity is, supposedly, restored at $x=1$.
It seems that we can then consider only the models which associate the
(super)conductivity suppression with a transfer of holes to an itinerant,
but different from the undoped YBCO, state, which should furthermore be
prone to localization. 
At first glance the only theoretical 
model that satisfy this criterion is  the itinerant model of Liechtenstein
and Mazin\cite{my,my1}. I
will show below that contrary to the claim in the Fehrenbacher and Rice paper%
\cite{FR} (FR), their model is {\it also} compatible with  metallic and
possibly superconducting behavior in PrBa$_{2}$Cu$_{3}$O$_{7}.$\cite{other}
 In fact, it
turns out that the difference between the ``local'' FR model and the
``itinerant'' LM model is much smaller than it was thought to be; if handled
correctly, the FR model never renders localized states. This new
understanding means that the physics of  superconductivity suppression
and its possible recovery is essentially the same in both model. There is
still a quantitative difference between the two, which is hard to access
experimentally, but which is now of limited importance. A generic model that
can be nicknamed FRLM explains the {\it entire body }of existing experimental
results and does not seem to have any sensible alternative. An exciting fact 
is that this generic model not only provides a possibility for metallic and
superconducting behavior of PrBa$_{2}$Cu$_{3}$O$_{7}$, but that it also
predicts the superconducting holes in it to be of entirely different
physical nature than the carrier in familiar high-$T_c$ superconductors. 

Let me start with a brief reminder of the essence of the FR model. The crystal
structure of the YBCO family cuprates is such that a rare earth ion and four
nearest oxygens form a nearly perfect cube. Moreover, among 7 orbitals of $f$%
-symmetry\label{a} there is one, $xyz,$ which has eight equivalent lobes
directed along eight directions [$\pm 1,\pm 1,\pm 1$]. In the standard
coordinate system where $x,$ $y$ correspond to the CuO bond directions the
same orbital is $(x^{2}-y^{2})z.$ Since this $f$-orbital extends directly
towards neighboring oxygens, one expects a noticeable $pf\sigma $ hopping
between $RE$ and O. Thus the electronic structure of a $RE$Cu$_{2}$O$_{4}$
bilayer breaks into two weakly interacting subsystems: usual Cu-O $pd\sigma $
bands, of which two antibonding ones cross the Fermi level in YBCO, and $RE$%
-O $fd\sigma $ states. Oxygen $p$ states directed along the Cu-O bonds (``$%
p\sigma $'' orbitals) participate in the former and those perpendicular to
the bonds (``$p\pi $'' orbitals) in the latter. If we start with a cluster
of one Pr and eight surrounding oxygens and considered formation of an
antibonding state of the $f_{(x^{2}-y^{2})z}$ Pr orbital and eight oxygen $p$
orbitals pointing directly towards Pr, we find one bonding, one antibonding,
and 6 non-bonding states. If the energy difference between the bare O $p$
level and bare Pr $f$ level is not too large, the energy of this antibonding
state, $pf\sigma ^{*},$ may become higher than that of the $pd\sigma ^{*}$
Cu-O state and will pull some holes out of the latter. Whether or not this
will happen depends on the $p-f$ energy separation and the $p-f$ hopping
integral. Suppression of superconductivity in $RE_{1-x}$Pr$_{x}$Ba$_{2}$Cu$%
_{3}$O$_{7}$ is thus ascribed to the hole transfer from the superconducting $%
pd\sigma ^{*}$ band into the $pf\sigma ^{*}$ state. An indispensable
component of this model is localization of carriers promoted into the $%
pf\sigma ^{*}$ state. FR\cite{FR} argued that the oxygen orbitals forming
this state form the 45$^{\circ }$ angle with the CuO planes and thus the
orbitals of the same oxygen pointing towards neighboring Pr ions are
orthogonal to each other. They also neglected direct hopping between the
oxygen $p\pi $ orbitals. In such an approximation the effective bandwidth of
the $pf\sigma ^{*}$ band is zero, and the holes there are localized by
infinitesimally small disorder. This was the original explanation of the lack
of (super)conductivity in PrBa$_{2}$Cu$_{3}$O$_{7}.$ Furthermore, since this
model renders noticeable presence of Pr states at the Fermi energy, one
expects the Curie temperature for ordering of the Pr moments to be much
higher than for other $RE$Ba$_{2}$Cu$_{3}$O$_{7},$ which is indeed the case.
A drawback is that total localization in the FR model does not let 
the given PrO$%
_{8}$ cluster be influenced in any direct way by the rare earths filling
other cells, in contradiction with the experiment: the $T_{c}$ suppression
rate even at low doping depends strongly on the host rare earth.

The LM model\cite{my} differs from the FR model in essentially only one
aspect: direct hopping between oxygen orbitals\cite{note} is taken into
account. LM calculated this hopping as well as other relevant parameters of
the electronic structure numerically using LDA+U method including Coulomb
correlation in the rare earth $f$-shell. They found substantial O-O hopping
so that independent of the value of the $pf\sigma $ hopping (and even the
presence of the $f$ orbital) the FR states were forming a dispersive band.
It was originally thought\cite{my}
 that an advantage of the LM model over the FR model was that a
dispersive band would hybridize with all rare earth ions in the crystal, and
its position {\it before} doping with Pr would depend on the position of the 
$f$-level in the host rare earth. This naturally, and with reasonable
quantitative agreement, explains the different rates of suppression with
different host $RE$\cite{my}. This model was however criticized\cite{FR1}
because localization of carriers in such a dispersive band requires finite
disorder and one expects stoichiometric PrBa$_{2}$Cu$_{3}$O$_{7}$ to be
metallic. Another prediction which was seemingly different from that of the
FR model was that the holes transferred to the $pf\sigma ^{*}$ band
concentrate near the center of the Brillouin zone; that is, near {\bf k}$%
=(\pi /a,\pi /b).$ This prediction could be indirectly checked by measuring
the ratio of the out-of-plane ($p_{z})$ and in-plane ($p_{x,y})$ oxygen
characters. At least qualitatively, this prediction was confirmed by the
experiment\cite{merz}.

Interestingly, it was not noticed until very recently\cite{my1} that the
geometric argument of FR\cite{FR} was incorrect: in fact, the angle that an
O-Pr bond forms with the CuO$_{2}$ planes is not 45$^{\circ },$ but $\tan
^{-1}(1/\sqrt{2})\approx 35^{\circ }16^{\prime },$ which means that even in
the FR limit of no direct O-O hopping, the $pf\sigma ^{*}$ states form a band
whose dispersion is defined by the Pr-O hopping amplitude $t_{pf\sigma }.$
Below I show how this band forms, using the nearest-neighbor tight-binding
Hamiltonian.

Let us begin with some notations: first, we neglect the (very small) $z$%
-dispersion. (This means that all orbitals we consider are antisymmetric
with respect to $z\rightarrow -z$ reflection, like the $f_{z(x^{2}-y^{2})}$
orbital.) Then the two plane problem is equivalent to a single plane. If
we include all nearest neighbors, the following orbitals contribute to the
FR band: (1) Pr $z(x^{2}-y^{2}),$ (2) O2 $z,$ (3) O3 $z,$ (4) O2 $y,$ (5) O3 
$x,$ (6) Cu $xy,$ (7) Cu $yz$, and (8) Cu $zx.$ Their in-plane 2D symmetries
are, respectively, $x^{2}-y^{2},$ $s,$ $s,$ $y,$ $x,$ $xy,$ $y,$ and $x,$
which simplifies the task of the tight-binding description of the band
structure. Let us now identify the largest hopping amplitudes between these
orbitals. According to FR, this is $pf\sigma $, which we shall denote $%
t_{pf}.$ It controls the following hoppings: $t_{12}=t_{13}=\sqrt{\frac{5}{27%
}}t_{pf},$ $t_{14}=t_{15}=\sqrt{\frac{10}{27}}t_{pf}=\sqrt{2}t_{12}.$ This
parameter defines the effect of the rare earth substitution on the FR band.
Another hopping, which is the strongest according to LM, is of $pd\pi $
type, denoted $t_{pd}.$ The hopping amplitudes $t_{28,}$ $t_{37,}$ $t_{46},$
and $t_{56}$ all are equal to $t_{pd}.$ This parameter defines the
dispersion of the FR band in the absence of the $f$ states; {\it e.g.}, in
YBa$_{2}$Cu$_{3}$O$_{7},$ or in the spin-minority channel of  PrBa$_{2}$Cu$%
_{3}$O$_{7}.$ Let us first consider these two hoppings separately.

\begin{figure}[tbp]
\centerline{\epsfig{file=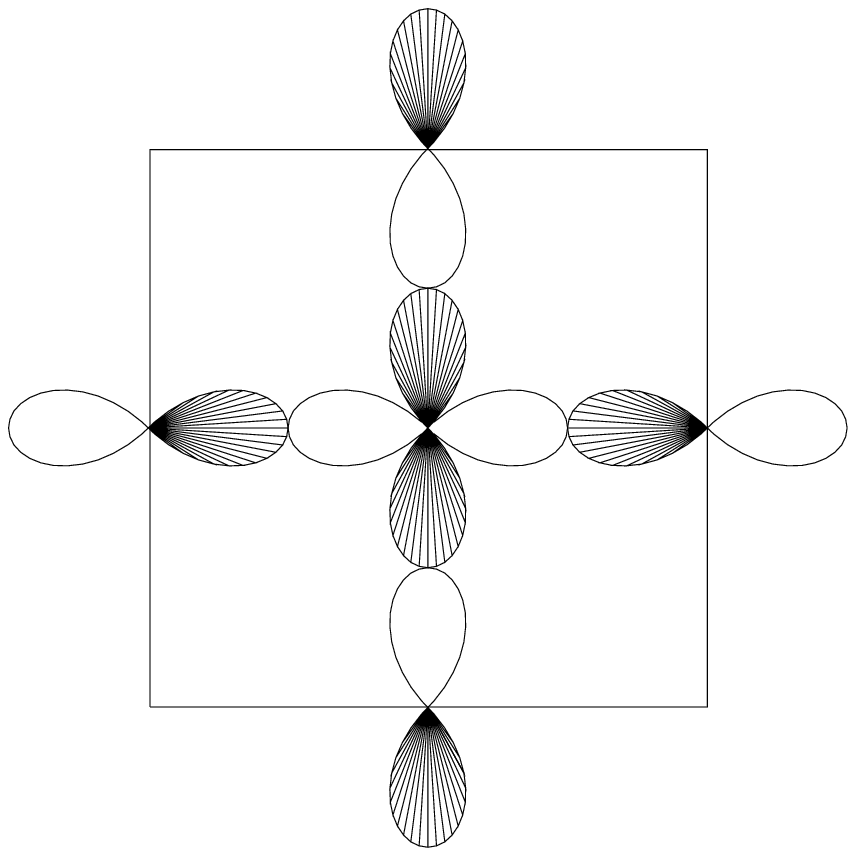,width=0.8\linewidth}}
\centerline{\epsfig{file=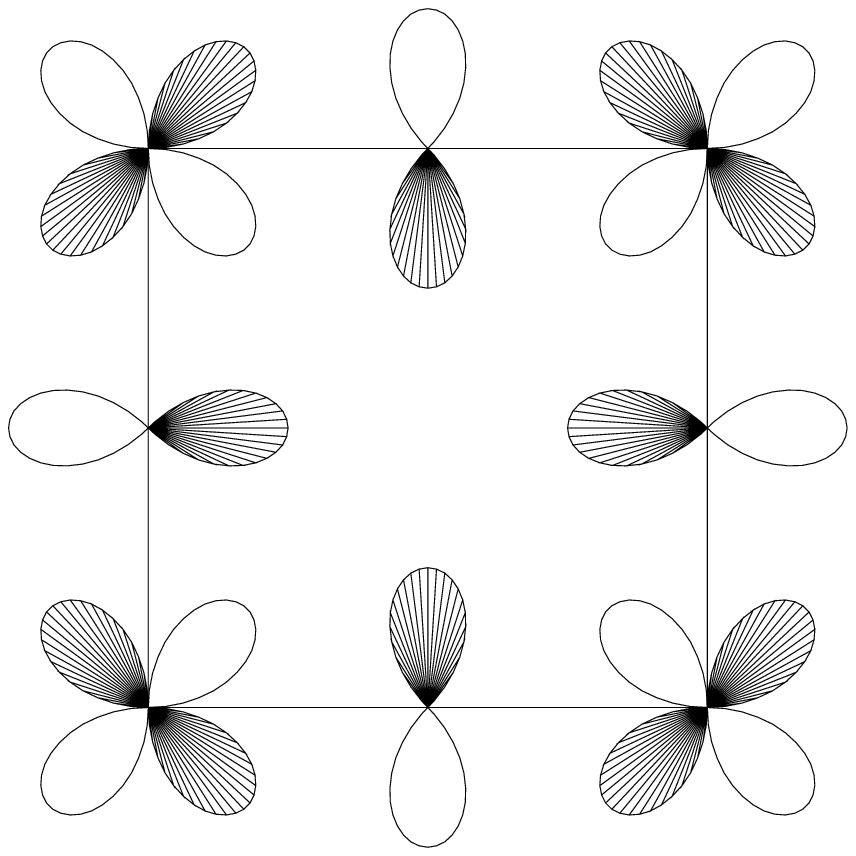,width=0.8\linewidth}}
\centerline{\epsfig{file=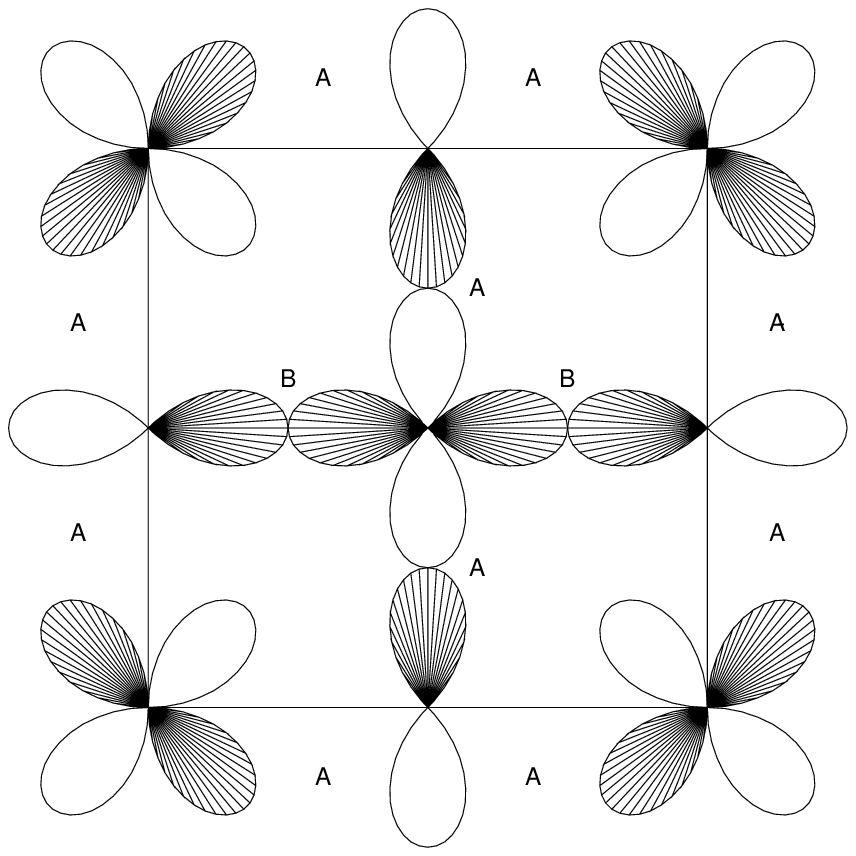,width=0.8\linewidth}}
\vspace{0.1in} \setlength{\columnwidth}{\linewidth} \nopagebreak
\caption{Tight-binding orbitals at the point {\bf S}=($\pi,\pi$) projected
onto $x-y$ plane. Upper panel: FR model, $t_{pd}=0$. The Pr ion
is in the center.
Middle panel: LM model for YBa$_2$Cu$_3$O$_7$, no Pr $z(x^2-y^2)$ orbital,
finite $t_{pd}$. Cu2 ions are in the corners. Lower panel: Illustration of
inability of the O2 $y$ and O3 $x$ orbitals to make an antibonding
combination simultaneously with the Cu2 $xy$ orbital (in the corners) and Pr 
$f$ orbital (in the center). }
\label{tb4}
\end{figure}

The FR model corresponds to an approximation $t_{pd}=0$. Dispersion of the
oxygen $p\pi $ states is completely neglected; an isolated Pr impurity forms
a localized antibonding state, shifted up with respect to the bare O $p$
level by 
\begin{equation}
\epsilon _{{\bf k}}-E_{p}=\delta \epsilon =\frac{5}{9}\frac{8t_{pf}^{2}}{%
E_{p}-E_{f}},
\end{equation}
where 8 stands for the eight neighboring oxygens. It is assumed that $%
t_{pf}\ll $ $E_{p}-E_{f}.$ In the opposite limit, when all rare earth sites
are occupied by Pr, a narrow band is formed with the dispersion 
\begin{equation}
\epsilon _{{\bf k}}-E_{p}=\delta \epsilon -\delta \epsilon \cos 2{\varphi}
(\cos ak_{x}+\cos bk_{y})/2,  \label{FR}
\end{equation}
where $\varphi =\arctan (1/\sqrt{2})$ is the angle that the Pr-O bond forms
with the $xy$ plane. Had this angle been 45$^{\circ },$ as assumed by FR,
the band would be dispersionless and thus fully localized. In reality, it
should acquire a finite bandwidth $W=\delta \epsilon \cos 2\varphi =\delta
\epsilon /3$, even were without Pr. This is an example of
dispersion due to nonorthogonality: the Hamiltonian written in terms of the
oxygen orbitals pointing towards Pr is diagonal, but such a basis is
nonorthogonal and that results in dispersion. Note that the top of the band
occurs at the ($\pi ,\pi )$ point and that is where the holes go from the $%
pd\sigma $ superconducting band. Fig.\ref{tb4} illustrates that indeed at
this point the $pf\sigma $ interaction is antibonding along all bonds.

Now we consider the case of finite $t_{pd}$ and no $f$ states. For
simplicity, we let the energy of the Cu $d$ level be the same as the energy
of the O $p$ level. Then four O $p$ orbitals and three Cu $d$ orbitals form
three antibonding bands (besides the bonding and nonbonding bands): 
\begin{eqnarray}
\epsilon _{{\bf k}}-E_{p} &=&2t_{pd}\sin \frac{ak_{x}}{2} \\
\epsilon _{{\bf k}}-E_{p} &=&2t_{pd}\sin \frac{bk_{y}}{2} \\
\epsilon _{{\bf k}}-E_{p} &=&2t_{pd}\sqrt{\sin ^{2}\frac{ak_{x}}{2}+\sin ^{2}%
\frac{bk_{y}}{2}}.  \label{LM}
\end{eqnarray}
The top of the highest (third) band is again at ($\pi ,\pi ),$ as
illustrated in Fig. \ref{tb4}, showing again the antibonding interactions
for all bonds.

The case of $pd\pi $ and $pf\sigma $ interactions taken
together cannot be solved
analytically. Before reporting the numerical results, we make one additional
observation: since both cases separately produce dispersive bands with the
maximum at ($\pi ,\pi ),$ one might expect this effect (band dispersion) to
be enhanced when both interaction are included. One can easily see (Fig.\ref
{tb4}), that this is not true: the configuration of the O $p$ orbitals,
which is antibonding in the first case, is nonbonding in the second case,
and vice versa.

The main shortcoming of the original FR model was its inability to describe
the different rate of $T_{c}$ suppression with the different rare earth
hosts. The LM model with its dispersive $pf\sigma $ provides a natural
explanation. However, we can see comparing Eqs.(\ref{FR}) and (\ref{LM}),
 that at
low doping, that is near the ($\pi ,\pi )$ point, the shape of the FR band
is very similar to that of the LM band. Note that the scale of the
dispersion, $i.e.$ the effective masses, may be different --- the FR band
should be heavier than the LM band and thus easier to localize. However, for
the ideal stoichiometric  PrBa$_{2}$Cu$_{3}$O$_{7}$ at zero temperature both
models give a metal and possibly a superconductor.

Another indirect argument in favor of the LM model over the FR model was
deduced from recent near-edge X-ray absorption experiments\cite{merz}. It
was found that O $p_{\pi }$ orbitals form a relatively small
angle with the CuO$_{2}$ planes: 20 to 25$^{\circ }$ at the doping level of
80\% Pr. This is closer to the prediction of the LM model\cite{my1} (15 to 18%
$^{\circ })$ than to the original FM prediction\cite{FR} (45$%
^{\circ }),$ and even to the corrected number of 36$^{\circ }.$ However, the
fact that the correct FR model is nonorthogonal not only yields a finite
bandwidth in pure  PrBa$_{2}$Cu$_{3}$O$_{7},$ but also makes the average
angle of the O $p_{\pi }$ orbitals dependent on doping at small doping ---
similar, but quantitatively different from the prediction of the LM model.

Indeed, the orbitals around an isolated Pr impurity are tilted by $\varphi
=\arctan (1/\sqrt{2});$ it is however obvious that if an oxygen atom has Pr
ions on {\it both} sides, the highest state does not include O($p_{z})$
character of this oxygen at all. In the low doping limit, for the Pr
concentration $x,$  the probability for an oxygen to have 1 Pr neighbor is $%
\nu _{1}=2x(1-x),$ and to have 2 neighbors is $\nu _{2}=x^{2}.$ Thus the
average $p_{z}$ character for the in-plane oxygen holes, seen in an
experiment like Ref.\cite{merz} is $n_{z}=\nu _{1}\sin ^{2}\varphi ,$ while
the total number of the holes in the FR state is $n=\nu _{1}+\nu _{2}$. The
average tilting angle is thus $\sin ^{2}\alpha =\frac{2}{3}\frac{1-x}{2-x}$%
. For $x=0.8$ we find $\alpha \approx 19.5^{\circ },$ in excellent agreement
with the experiment. Moreover, this number is the lower bound on $\alpha $
in the FR model, because at large $x$ one cannot neglect dispersion of the
FR band, which will force $\alpha $ to deviate from the formula above (at $%
x=1$ the FR model should give the same number as the LM model, which is\cite
{my1} about 20$^{\circ }),$ so that at $x=0.8$ we find $\alpha \agt20^{\circ
}.$ Interestingly, while both the FR  and LM models predict a dependence of
the angle on concentration, and both must give the same value at $x=1$ ($%
\alpha $ does not depend on the effective mass at this point), they predict
the opposite dependences: in the LM model $\alpha $ falls to zero when $%
x\rightarrow 0,$ while in the FR model it increases up to $\alpha =$ $%
\varphi \approx 36^{\circ }.$ 

My conclusion is that after being corrected to take into account the right
geometry of the Pr-O bonds, the FR model provides better agreement with the
experiment than the LM model. I emphasize that after such correction the
difference between the two models is not the difference between a band model
and a localized model, but the difference between two band models, one where
dispersion originates from the $pf\sigma $ Pr-O hopping, and another where
it appears mostly due to the $pd\pi $ Cu-O hopping. The fact that the former
appears more successful (however, the final word will be said by an
experiment accessing $x$ dependence of the angle $\alpha ),$ does not mean
that the $pd\pi $ hopping is negligibly small. As illustrated on Fig.1c, it
merely means that the dispersion due to this hopping is weaker than that due
to $t_{pf\sigma }.$ 

To summarize, current experimental situation in Pr$_{x}RE_{1-x}$Ba$_{2}$Cu$%
_{3}$O$_{7}$ is such that the band version of the
Fehrenbacher-Rice model presented
here explains all existing experiments addressing
superconducting and transport properties of this system, including the
recent observation of superconductivity at full substitution. In fact,  PrBa$%
_{2}$Cu$_{3}$O$_{7}$ is a more novel superconductor than all other cuprate
high $T_{c}$ materials known: it is the only one where superconducting
carriers are not residing in the Cu$(x^{2}-y^{2})-$O$(p_{\sigma })$ bands,
but are of entirely different character.

\end{multicols}


\begin{references}
\bibitem{other} I concentrate in this Letter on these two models\cite{my,FR}
because no other model I am aware of has any mechanism for higher 
conductivity and, eventually, superconductivity in fully substituted
PrBa$_2$Cu$_3$O$_7$ compared with a partially substituted Pr$_x$Y$
_{1-x}$Ba$_2$Cu$_3$O$_7$.
\bibitem{note}  In this context ``direct hopping'' includes also all
assisted hoppings which do not involve the Pr $f$ states, $e.g.$ $via$ Cu $%
d_{xz}$ and $d_{yz}$ orbitals.
\end{references}
\end{document}